\documentclass[11pt,a4paper]{article}
\pdfoutput=1
\usepackage{jheppub,yfonts}

\usepackage{amsmath}
\usepackage{amssymb}
\usepackage{amsthm}
\usepackage{mathrsfs}
\usepackage{graphicx}
\usepackage{fancyhdr}
\usepackage{array}
\usepackage{latexsym}
\usepackage[all]{xy}
\usepackage{eufrak}
\usepackage{euscript}
\usepackage{enumerate}
\usepackage{dsfont}
\usepackage{slashed}
\usepackage{hyperref}
\usepackage{caption}

\newcommand{\be}{\begin{equation}}
\newcommand{\ee}{\end{equation}}
\newcommand{\bea}{\begin{eqnarray}}
\newcommand{\eea}{\end{eqnarray}}

\def \v {\vec}
\def \mw {\mathfrak{w}}

\title{Vanishing DC holographic conductivity from a magnetic monopole condensate}

\author[a]{Romulo Rougemont,}
\author[a,b]{Jorge Noronha,}
\author[c]{Carlos A. D. Zarro,}
\author[c]{Clovis Wotzasek,}
\author[d]{Marcelo S. Guimaraes,}
\author[d,e]{Diego R. Granado}

\affiliation[a]{Instituto de F\'{i}sica, Universidade de S\~{a}o Paulo, C.P. 66318, 05315-970, S\~{a}o Paulo, SP, Brazil}
\affiliation[b]{Department of Physics, Columbia University, 538 West 120th Street, New York, NY 10027, USA}
\affiliation[c]{Instituto de F\'{i}sica, Universidade Federal do Rio de Janeiro, 21941-972, Rio de Janeiro, RJ, Brazil}
\affiliation[d]{Instituto de F\'{i}sica, Universidade do Estado do Rio de Janeiro, 20550-013, Rio de Janeiro, RJ, Brazil}
\affiliation[e]{Ghent University, Department of Physics and Astronomy, Krijgslaan 281-S9, 9000 Gent, Belgium}

\date{\today}

\abstract{We show how to obtain a vanishing DC conductivity in 3-dimensional strongly coupled QFT's using a massive 2-form field in the bulk that satisfies a special kind of boundary condition. The real and imaginary parts of the AC conductivity are evaluated in this holographic setup and we show that the DC conductivity identically vanishes even for an arbitrarily small (though nonzero) value of the 2-form mass in the bulk. We identify the bulk action of the massive 2-form with an effective theory describing a phase in which magnetic monopoles have condensed in the bulk. Our results indicate that a condensate of magnetic monopoles in a 4-dimensional bulk leads to a vanishing DC holographic conductivity in 3-dimensional strongly coupled QFT's.}
   
\keywords{Holography, AC conductivity, magnetic monopole, condensate.}

\emailAdd{romulo@if.usp.br, noronha@if.usp.br, carlos.zarro@if.ufrj.br, clovis@if.ufrj.br, msguimaraes@uerj.br, diegorochagrana@uerj.br}
 
\begin{document}

\maketitle

\setlength{\parskip}{8pt}


\section{Introduction}
\label{sec1}

The anti-de Sitter/conformal field theory (AdS/CFT) correspondence or gauge/string duality \cite{adscft1,adscft2,adscft3} is one of the major breakthroughs which emerged from string theory in the last years. In its classical (super)gravity limit it becomes a correspondence between classical (super)gravity in asymptotically AdS spaces and strongly coupled quantum field theories (QFT's) with a large rank non-Abelian gauge group defined on the conformally flat boundary of such spaces. Thus, the holographic gauge/gravity correspondence maps difficult non-perturbative phenomena of strongly interacting systems into manageable classical gravity setups in higher dimensions. 

One of the many attractive features of the holographic correspondence is the fact that it can be used to compute non-equilibrium transport properties in strongly-coupled theories with gravity duals. This has been extensively investigated in the last decade in the context of the strongly-coupled quark gluon plasma \cite{Kovtun:2004de,CasalderreySolana:2011us,Finazzo:2014cna}. Recently, the holographic duality has been also used in the description of some properties of strongly correlated condensed matter systems, as reviewed in \cite{adscmt1,adscmt2,adscmt3,adscmt4,adscmt5,adscmt6}.

In the present work we focus on a specific transport coefficient, namely, the AC conductivity. For definiteness, we consider here a  $(2+1)$-dimensional QFT defined at the boundary of a $(3+1)$-dimensional asymptotically AdS background. In this case, it has been shown in \cite{sachdev,herzog,iqbal-liu} that the AC conductivity associated with a conserved vector current sourced by the boundary value of an Abelian 1-form bulk gauge field in an AdS$_{3+1}$-Schwarzschild background displays no dependence on the frequency of the externally applied electric field at the boundary. This remarkable result was linked in \cite{sachdev} with the fact that Einstein-Maxwell theory's in a $(3+1)$-dimensional bulk is electromagnetically self-dual. This observation was later used in \cite{sachdev-conductivity} to present a way of turning on a nontrivial frequency-dependence for the QFT's conductivity by breaking this bulk electromagnetic self-duality via the introduction of higher derivative interactions in the bulk action for the metric and the Maxwell field.

In this paper we introduce a new way of breaking the electromagnetic self-duality of Einstein-Maxwell's theory in a $(3+1)$-dimensional bulk which induces a nontrivial frequency-dependence for the AC conductivity of the strongly coupled QFT in $(2+1)$-dimensions. This is done in Section \ref{sec2} by sourcing the QFT conserved 1-form current with the boundary value of a massive 2-form bulk field satisfying a special boundary condition, namely, that the effective mass of the 2-form bulk field vanishes at the boundary. The requirement of finiteness of the action imposes that the boundary value of this 2-form field reduces to the exterior derivative of a 1-form Maxwell gauge field, which then sources the boundary QFT conserved vector current operator. Remarkably, in such a scenario the QFT DC conductivity vanishes, while at high frequencies one recovers the constant result for the conductivity characteristic of Einstein-Maxwell's theory. These results display a certain degree of universality in the sense that they are valid for any isotropic black brane background. In Section \ref{sec2.3}, in order to evaluate numerically the AC conductivity for intermediate values of frequency, we use the AdS$_{3+1}$-Schwarzschild background and show that the frequency-dependent behavior of the AC conductivity depends on how the effective mass of the 2-form bulk field varies in the chosen bulk geometry.

The setup we discuss in this work constitutes a new way to obtain a vanishing DC holographic conductivity in strongly coupled QFT's with gravity duals defined in asymptotically AdS spacetimes in several dimensions. We further argue in Section \ref{sec3} that the massive 2-form bulk field may be linked to the Maxwell 1-form bulk field through a condensation of magnetic monopoles in the bulk. In fact, we argue that the vanishing DC holographic conductivity should be a generic property of 3-dimensional QFT's that can be described using a 4-dimensional effective action modeling a phase in which magnetic monopoles have condensed in the bulk. We present our conclusions and outlook in \ref{sec4}. We also discuss in Appendix \ref{apa} that a similar reasoning using a Proca field in the bulk, instead of a massive rank 2 tensor, leads to different results for the conductivity which are qualitatively the same as those obtained in the context of holographic superconductors in the probe approximation discussed in \cite{HHH}. 

Throughout this paper we use natural units $c=\hbar=k_B=1$ and a mostly plus metric signature. Greek indices are used to denote bulk coordinates while Latin indices denote coordinates parallel to the boundary of the asymptotically AdS space.

\section{Massive 2-form bulk field and holographic conductivity}
\label{sec2}

We begin by writing down a bulk action for a quadratic massive 2-form field, $K_{\mu\nu}$, with a specific interaction with a (dimensionless) real scalar field, $M$, in a curved 4-dimensional spacetime\footnote{Note that here we only consider the probe limit in which the (fixed) background metric influences the 2-form field and the scalar field but those do not backreact on the metric. A natural generalization to be pursued in the future consists in considering also the metric field to be dynamic.} $\mathcal{M}_{3+1}$, which we assume to be asymptotically AdS$_{3+1}$
\begin{align}
S_{\textrm{2-form+scalar}}=-\int_{\mathcal{M}_{3+1}}d^{4}x\sqrt{-g} &\left[\frac{1}{12M^2}\left(\nabla_\mu K_{\alpha\beta} +\nabla_\alpha K_{\beta\mu}+\nabla_\beta K_{\mu\alpha}\right)^2 +\frac{\Lambda^2}{4}K_{\mu\nu}^2+\right.\nonumber\\
&\left.+\frac{\Lambda^2}{2\bar{g}^2}\left(\nabla_\mu M\right)^2 +V(M)\right],
\label{diffeopreserving}
\end{align}
where $\Lambda$ is the mass of the 2-form field, $\bar{g}$ is the (dimensionless) coupling of the scalar field with the background, and $V(M)$ is an arbitrary potential for the scalar field. The above action is discussed, for instance, in Section 3.8 of \cite{ripka} in the context of the dual Abelian Higgs model describing a monopole condensate. We shall come back to this point in Section \ref{sec3} but for now let us discuss how we may employ \eqref{diffeopreserving} in a very simple manner to obtain new results concerning the holographic conductivity associated to a conserved vector current in the boundary QFT.

Let us first rescale $K_{\mu\nu}\mapsto K_{\mu\nu}/\Lambda$ in \eqref{diffeopreserving} and define also the rescaled scalar field
\begin{align}
m(u):=\Lambda\, M(u),
\label{2.2}
\end{align}
where we assumed that the scalar field depends only on the holographic radial coordinate, $u$. Then, the sector of the action \eqref{diffeopreserving} which depends on the 2-form field may be written as
\begin{align}
S=-\frac{1}{4}\int_{\mathcal{M}_{3+1}}d^{4}x\sqrt{-g}\left[\frac{1}{3m^2(u)}\left(\nabla_\mu K_{\alpha\beta} +\nabla_\alpha K_{\beta\mu}+\nabla_\beta K_{\mu\alpha}\right)^2 +K_{\mu\nu}^2\right].
\label{2.1}
\end{align}
The action \eqref{2.1} may be loosely thought as the action for a quadratic massive 2-form field with an effective radial-dependent mass, $m(u)$. However, this action just by itself explicitly violates diffeomorphism invariance since its energy-momentum tensor is not covariantly conserved. This is simply due to the fact that the scalar field, $m(u)$, which couples to the 2-form field, is a dynamic field and the partial action \eqref{2.1} does not take into account the dynamics of this scalar field, which is considered in this sense as an external field. 

On the other hand, the complete action \eqref{diffeopreserving} (which includes the dynamics of the scalar field/effective mass) is diffeomorphism-invariant and has a convariantly conserved energy-momentum tensor. Nonetheless, the partial action \eqref{2.1} does provide a shortcut for the calculation of 2-point correlation functions of components of the 2-form field. Note that the two terms present in the dynamical sector for the effective mass in the complete action \eqref{diffeopreserving}, which are being neglected in the simpler partial action \eqref{2.1}, do not couple directly to the 2-form field. Therefore, the equations of motion and also the functional form of the 2-point correlation functions of components of the 2-form field are exactly the same when derived from the complete action \eqref{diffeopreserving} or from the partial action \eqref{2.1}. Consequently, in what concerns the calculation of 2-point correlation functions of components of the 2-form field, the neglected dynamics of the effective mass in the partial action \eqref{2.1} only influences the results indirectly by restricting the possible forms of the effective mass given a certain choice for the scalar field potential in \eqref{diffeopreserving}. However, since the potential for the effective mass was taken to be arbitrary at this point, one can take a prescribed profile for $m(u)$ in the partial action \eqref{2.1}, calculate the 2-point correlation functions for components of the 2-form field using this simpler action, and the corresponding results may be in principle equivalently obtained from the complete action \eqref{diffeopreserving} by choosing a potential which gives the used profile for $m(u)$ as a solution of the classical equations of motion for the scalar field. In this way, the partial action \eqref{2.1} suffices to capture the essential physics required for the calculation of 2-point correlation functions of components of the 2-form field and, thus, it constitutes a much simpler path than the one where the complete action \eqref{diffeopreserving} is used from the outset. Therefore, in the following we shall only consider the partial action in \eqref{2.1}.

The holographic dictionary \cite{adscft1,adscft2,adscft3} states that the boundary value of a massless 1-form bulk gauge field acts as a source for a conserved 1-form current at the boundary QFT. However, here we want instead the boundary value of the massive 2-form bulk field $K_{\mu\nu}$ in \eqref{2.1} (or, more generally, in \eqref{diffeopreserving}) to be the source for a conserved vector current at the boundary. Indeed, the essential observation is that while a conserved 1-form current couples to a massless 1-form gauge field at the boundary, this does not imply that one must necessarily have a massless 1-form gauge field in the bulk. Another possible way to obtain a massless 1-form gauge field at the boundary sourcing a conserved vector current operator, which shall be pursued in this work, consists in having a massive 2-form field in the bulk with an effective radial-dependent mass that vanishes at the boundary. In this case, the requirement of finiteness for the action \eqref{2.1} implies the following \textit{boundary condition} for the 2-form field\footnote{The superscript 0 denotes the boundary value of the field.}
\begin{align}
m(u)\biggr|_{\partial\mathcal{M}_{3+1}}\rightarrow 0 \Rightarrow K_{\mu\nu}\biggr|_{\partial\mathcal{M}_{3+1}}\rightarrow K_{ij}^0=\partial_{[i}A_{j]}^0\,.
\label{2.3}
\end{align}
Therefore, as one approaches the boundary the numerator of the kinetic term for $K_{\mu\nu}$ in \eqref{2.1} vanishes and the mass term for the 2-form field reduces to the kinetic term for a boundary Maxwell field\footnote{Thus, one should expect that at high frequencies the conductivity calculated via the action \eqref{2.1} agrees with the one calculated via the Maxwell action in the bulk. This is indeed the case as we shall show in the following.}. Note that the same reasoning implies that if we take the mass scale $\Lambda$ in \eqref{2.2} to zero we obtain from \eqref{2.1} a Maxwell action in the bulk. This means that $\Lambda$ gives a connection between a bulk Maxwell action and the action for a massive 2-form field, whose physical meaning we shall elaborate upon in Section \ref{sec3}. It is also important to observe that imposing that the effective mass function for the 2-form field in the partial action \eqref{2.1} vanishes at the boundary, from the point of view of the complete action \eqref{diffeopreserving}, corresponds to take this effective mass as a relevant deformation of the boundary QFT, which modifies only the infrared physics corresponding holographically to the interior of the bulk. Furthermore, besides the requirement that the effective mass function vanishes at the boundary, we shall also assume that it is regular in the interior of the bulk.

It is useful at this point to contrast our setup with the one used in the calculation of the AC conductivity in Einstein-Maxwell's theory in an AdS$_{3+1}$-Schwarzschild black brane background \cite{sachdev}. In this case, one has the Maxwell action in the bulk
\begin{equation}
S_{\textrm{Maxwell}}=-\frac{1}{4}\int_{\rm{AdS}_{3+1}}d^{4}x\sqrt{-g}\,\mathcal{F}_{\mu\nu}^2\,,
\end{equation}
where $\mathcal{F}_{\mu\nu}=\nabla_\mu A_\nu-\nabla_\nu A_\mu$. The AC conductivity associated with the boundary value of the bulk Maxwell field is just a constant \cite{sachdev,herzog,iqbal-liu}\footnote{In these references the conductivity is given by $1/g^2$, where $g$ is the Maxwell gauge coupling. Since in four dimensions this coupling is just a dimensionless constant, one may trivially absorb it in the normalization of the Maxwell field and the corresponding AC conductivity is set to 1, as in \eqref{2.4}. If one wants to keep this coupling in the expression for the Maxwell AC conductivity one should then correspondingly insert a constant factor of $g^2$ in the denominator of the mass term in the action \eqref{2.1}, since, as discussed before, \eqref{2.3} implies that this mass term goes to the Maxwell action as one approaches the boundary.}
\begin{align}
\sigma_{\textrm{Maxwell}}(\omega)=1.
\label{2.4}
\end{align}
The DC conductivity is given by the real part of the zero frequency limit of the AC conductivity and, since the Maxwell AC conductivity is a real-valued constant, the DC conductivity coincides with it.

\subsection{Boundary current propagator and conductivity from the on-shell action}
\label{sec2.1}

Now let us see how the conductivity behaves if one follows the alternative holographic approach discussed in the previous section based on the action \eqref{2.1} for a massive 2-form field satisfying the boundary condition \eqref{2.3}. The results of this section hold for any diagonal and isotropic black brane background of the form\footnote{We define $g_{tt}>0$.}
\begin{align}
g_{\mu\nu}=\textrm{diag}\left[g_{uu}(u),-g_{tt}(u),g_{xx}(u),g_{xx}(u)\right],\,\,\,g_{tt}(u)=H(u)(u_H-u),\,\,\, g_{uu}(u)=\frac{G(u)}{u_H-u},
\label{2.5}
\end{align}
where the boundary is located at $u=0$ while the horizon is at $u=u_H$ where $g_{tt}$ has a simple zero and $H(u_H)$ and $G(u_H)$ are finite. The Hawking temperature is given by
\begin{align}
T=\frac{\sqrt{g_{tt}'\,g^{uu}\,'}}{4\pi}\Biggr|_{u=u_H}=\frac{1}{4\pi}\sqrt{\frac{H(u_H)}{G(u_H)}}\,,
\label{2.6}
\end{align}
where the prime denotes a derivative with respect to the holographic coordinate, $u$.

As discussed for instance in \cite{iqbal-liu} the AC conductivity is given by
\begin{align}
\sigma_{ab}(\omega)=\frac{iG_{ab}^R(\omega)}{\omega};\,\,a,b=x,y,
\label{2.7}
\end{align}
where $G_{ab}^R(\omega)\equiv G_{ab}^R(\omega,\v{q}=\v{0})$ is the retarded thermal 2-point correlation function of the boundary QFT vector current evaluated at vanishing spatial momentum. Assuming spatial isotropy it follows that $\sigma_{xx}(\omega)=\sigma_{yy}(\omega)\equiv\sigma(\omega)$ \footnote{The off-diagonal Hall conductivity, $\sigma_{xy}(\omega)$, vanishes for the holographic setup considered here due to parity conservation. The Hall conductivity can be studied within our approach by including a topological term of the type $\sim \theta \,K_{\mu\nu}\varepsilon^{\mu\nu\lambda\rho}K_{\lambda\rho}$ in the Lagrangian displayed in \eqref{2.1}.}. In order to calculate this retarded propagator we follow the holographic prescription proposed in \cite{ss}, which was further justified and generalized in \cite{herzog-son,see-apdx-c,full-prescription}.

The equations of motion following from \eqref{2.1} read
\begin{align}
\partial_\mu\left[\frac{\sqrt{-g}}{m^2(u)}g^{\mu\rho}g^{\nu\sigma} g^{\alpha\beta}(\partial_\rho K_{\sigma\beta}+ \partial_\sigma K_{\beta\rho}+\partial_\beta K_{\rho\sigma})\right] -\sqrt{-g}g^{\nu\sigma}g^{\alpha\beta}K_{\sigma\beta}=0.
\label{4.1}
\end{align}
Let us now define the Fourier representation\footnote{For the sake of notation simplicity we distinguish a function from its Fourier transform only by their arguments.}
\begin{align}
K_{\mu\nu}(u,t,\v{x})=\int \frac{d\omega d^2\v{q}}{(2\pi)^3}e^{i(-\omega t+\v{q}\cdot\v{x})}K_{\mu\nu}(u,\omega,\v{q}).
\label{3.5}
\end{align}
After taking the limit of zero spatial momentum and substituting the Fourier mode $K_{\mu\nu}(u,t,\omega)\equiv e^{-i\omega t}K_{\mu\nu}(u,\omega)$ into (\ref{4.1}), we obtain
\begin{align}
K_{xy}''-\left(\frac{g_{xx}'}{g_{xx}}+\frac{g_{uu}'}{2g_{uu}}-\frac{g_{tt}'}{2g_{tt}}+ \frac{2m'(u)}{m(u)}\right)K_{xy}'+\frac{g_{uu}}{g_{tt}} (\omega^2-m^2(u) g_{tt})K_{xy}&=0,\label{4.2}\\
K_{tx}''-\left(\frac{g_{uu}'}{2g_{uu}}+\frac{g_{tt}'}{2g_{tt}}+\frac{2m'(u)}{m(u)}\right)K_{tx}' -m^2(u)g_{uu} K_{tx}+i\omega K_{ux}' &+\nonumber\\
-i\omega\left(\frac{g_{uu}'}{2g_{uu}}+\frac{g_{tt}'}{2g_{tt}}+\frac{2m'(u)}{m(u)}\right)K_{ux}&=0,\label{4.3}\\
K_{ux}-\frac{i\omega}{\omega^2-m^2(u) g_{tt}}K_{tx}'&=0,\label{4.4}\\
K_{ty}''-\left(\frac{g_{uu}'}{2g_{uu}}+\frac{g_{tt}'}{2g_{tt}}+\frac{2m'(u)}{m(u)}\right)K_{ty}' -m^2(u)g_{uu} K_{ty}+i\omega K_{uy}' &+\nonumber\\
-i\omega\left(\frac{g_{uu}'}{2g_{uu}}+\frac{g_{tt}'}{2g_{tt}}+\frac{2m'(u)}{m(u)}\right)K_{uy}&=0,\label{4.5}\\
K_{uy}-\frac{i\omega}{\omega^2-m^2(u) g_{tt}}K_{ty}'&=0,\label{4.6}\\
K_{ut}&=0\label{4.7}.
\end{align}

By substituting (\ref{4.4}) into (\ref{4.3}), we find the following decoupled equation for $K_{tx}$ (and also for $K_{ty}$)
\begin{align}
K_{tx}''+\left[\frac{g_{tt}'}{2g_{tt}}-\frac{g_{uu}'}{2g_{uu}}+ \frac{m^2(u) g_{tt}}{\omega^2-m^2(u)g_{tt}}\left( \frac{g_{tt}'}{g_{tt}}+\frac{2m'(u)}{m(u)} \right)\right]K_{tx}'+\frac{g_{uu}}{g_{tt}} (\omega^2-m^2(u) g_{tt})K_{tx}=0.
\label{4.8}
\end{align}
We can also recast (\ref{4.8}) into the following compact form by looking directly at the $tx$-component of the equation of motion (\ref{4.1}) in momentum space and using \eqref{4.4}
\begin{align}
\partial_u\left(\sqrt{\frac{g_{tt}}{g_{uu}}}\,\frac{K_{tx}'}{\omega^2-m^2(u) g_{tt}}\right)+\sqrt{\frac{g_{uu}}{g_{tt}}} K_{tx} = 0.
\label{4.9}
\end{align}

The asymptotic solution of \eqref{4.8} near the boundary behaves in the same way as the asymptotic solution for a Maxwell field\footnote{Here $u=\epsilon$ is an ultraviolet cutoff corresponding to a fixed $u$-slice of the asymptotically AdS$_{3+1}$ space near the boundary at $u=0$.}, $K_{tx}(u\rightarrow\epsilon,\omega)\sim C_1(\omega)+C_2(\omega)\epsilon$, as long as the mass function \eqref{2.2} near the boundary goes like $m(\epsilon)\sim\epsilon^a$, for $a=1$ or $a>3/2$. In these cases, one can impose the following Dirichlet boundary condition\footnote{Note that in \eqref{dirichlet} we are omitting the dependence of the Fourier modes on the spatial momentum since it does not contribute to $G_{xx}^R(\omega)$, which is the quantity we want to calculate.}
\begin{align}
\lim_{u\rightarrow 0} K_{tx}(u,t,\omega)=K_{tx}^0(t,\omega)=-i\omega A_x^0(t,\omega). 
\label{dirichlet}
\end{align}
In order to factor out the source term for the boundary QFT vector current operator, we define 
\begin{align}
K_{tx}(u,\omega)\equiv -i\omega A_x^0(\omega) F(u,\omega)
\label{ansatz}
\end{align}
and in terms of the function $F(u,\omega)$ the Dirichlet boundary condition \eqref{dirichlet} becomes $F(0,\omega)=1$.

In order to obtain the retarded propagator of the conserved vector current at the boundary one discards the piece of the total on-shell action evaluated at the horizon and take into account only the part evaluated at the boundary \cite{ss}. Furthermore, one must work with on-shell field configurations satisfying the in-falling wave condition at the horizon. Then, the sector of the on-shell boundary action contributing to $G_{xx}^R(\omega)$ is given by\footnote{We used the reality condition: $K_{tx}(x)=K_{tx}^*(x)\Rightarrow K_{tx}(u,-\omega,-\v{q})=K_{tx}^*(u,\omega,\v{q})$. Also, we employed the set of equations of motion \eqref{4.2} - \eqref{4.7} to set $K_{ut}=0$ and identify $K_{ux}$ as a function of $K'_{tx}$ in the on-shell boundary action.}
\begin{align}
S^{\textrm{boundary}}_{\textrm{on-shell}}=-\frac{1}{2}\int \frac{d\omega d^2\v{q}}{(2\pi)^3}A_x^0\,^*(\omega,\v{q})\left[-\lim_{\epsilon\rightarrow 0}\sqrt{\frac{g_{tt}(\epsilon)}{g_{uu}(\epsilon)}}\, \frac{\omega^2 F'(\epsilon,\omega,\v{q})}{\omega^2-m^2(\epsilon) g_{tt}(\epsilon)}\right]_{\textrm{on-shell}}^{\textrm{in-falling}} A_x^0(\omega,\v{q}) + (\cdots\,\!),
\label{4.12}
\end{align}
where $(\cdots\,\!)$ denotes terms that do not contribute to $G^R_{xx}(\omega)$.

As discussed before, for mass functions satisfying the near-boundary asymptotics $m(\epsilon)\sim\epsilon^a$, with $a=1$ or $a>3/2$, the asymptotic solution of the equation of motion for the 2-form field near the boundary is the same as the asymptotic solution for the Maxwell field, and one can easily show that the on-shell boundary action \eqref{4.12} remains finite in the limit $\epsilon\rightarrow 0$, such that one does not need to resort to the holographic renormalization procedure \cite{ren1,ren2,ren3,ren4,ren5} in these cases, i.e, the boundary condition for the massive 2-form ensures that the on-shell action is finite at the boundary just like the Maxwell action is in the same dimensionality. Then, we can immediately obtain from \eqref{4.12} and \eqref{2.7} the expressions for the retarded Green's function and the associated conductivity, respectively:
\begin{align}
G_{xx}^R(\omega)&=-\lim_{\epsilon\rightarrow 0} \sqrt{\frac{g_{tt}(\epsilon)}{g_{uu}(\epsilon)}}\, \frac{\omega^2 F'(\epsilon,\omega)}{\omega^2-m^2(\epsilon) g_{tt}(\epsilon)}\Biggr|_{\textrm{on-shell}}^{\textrm{in-falling}},\label{prop}\\
\sigma(\omega)&=\frac{iG_{xx}^R(\omega)}{\omega}=-\lim_{\epsilon\rightarrow 0} \sqrt{\frac{g_{tt}(\epsilon)}{g_{uu}(\epsilon)}}\, \frac{i\omega F'(\epsilon,\omega)}{\omega^2-m^2(\epsilon) g_{tt}(\epsilon)}\Biggr|_{\textrm{on-shell}}^{\textrm{in-falling}}.\label{cond}
\end{align}
It is important to observe here that \eqref{prop} corresponds to the 2-point correlation function of the same boundary QFT vector current operator as in the case of a Maxwell field in the bulk since, at the boundary, the source for this operator is exactly the same in both cases corresponding to a 1-form Abelian gauge field. Also, the dimension of this operator is exactly the same in both cases since as discussed before the asymptotic solution for the 2-form field near the boundary coincides with the asymptotic solution for the Maxwell field. Still, the results for the 2-point correlation functions in the cases of the 2-form field and the Maxwell field and, consequently their associated conductivities, are different in the infrared. As we shall discuss in Section \ref{sec3}, this is related to the fact that these two different pictures describe different phases of the system.

Note that in the deep ultraviolet, $\omega\gg\Lambda,\,T$, one can approximate $\omega^2-m^2(u) g_{tt}(u)\approx\omega^2$ and, thus, the decoupled equation of motion \eqref{4.8} for $F$ reduces to the equation of motion for the $x$ (or also the $y$) component of a Maxwell field in the general background \eqref{2.5}
\begin{align}
F''+\left(\frac{g'_{tt}}{2g_{tt}}-\frac{g'_{uu}}{2g_{uu}}\right)F'+\frac{\omega^2 g_{uu}}{g_{tt}}F=0,\,\,\,\textrm{for}\,\,\, \omega\gg\Lambda,\,T.
\label{2.24}
\end{align}
Therefore, in this ultraviolet regime, the Green's function in \eqref{prop} reduces to the Green's function associated with a Maxwell field on the background \eqref{2.5}
\begin{align}
G_{xx}^R(\omega)\biggr|_{\omega\gg\Lambda,\,T}=-\lim_{\epsilon\rightarrow 0} \sqrt{\frac{g_{tt}(\epsilon)}{g_{uu}(\epsilon)}}\, F'(\epsilon,\omega)\Biggr|_{\textrm{on-shell}}^{\textrm{in-falling}}.
\label{2.25}
\end{align}
Consequently, we necessarily recover the result for the Maxwell conductivity at high frequencies, as expected. This general analytic result will also be useful as a consistency check of the numerical results for the AC conductivity in Section \ref{sec2.3}.

In order to compute the AC conductivity \eqref{cond}, we substitute the Ansatz \eqref{ansatz} into \eqref{4.9} and define the following quantity
\begin{align}
\Pi(u,\omega)=-\sqrt{\frac{g_{tt}}{g_{uu}}}\, \frac{i\omega}{\omega^2-m^2(u) g_{tt}}\,\frac{F'(u,\omega)}{F(u,\omega)},
\label{Pi}
\end{align}
which obeys a first order ordinary differential equation \cite{iqbal-liu}
\begin{align}
\Pi'+i\sqrt{\frac{g_{uu}}{g_{tt}}}\left[\frac{(\omega^2-m^2(u)g_{tt})\Pi^2}{\omega}-\omega\right]=0.
\label{eqPi}
\end{align}
Using the Dirichlet boundary condition $\lim_{\epsilon\rightarrow 0} F(\epsilon,\omega)=1$, one finds that AC conductivity \eqref{cond} can be written in terms of $\Pi$ as follows
\begin{align}
\sigma(\omega)=\lim_{\epsilon\rightarrow 0}\Pi(\epsilon,\omega)\biggr|_{\textrm{on-shell}}^{\textrm{in-falling}}.
\label{condPi}
\end{align}

Since \eqref{eqPi} is a first order differential equation, one needs only one boundary condition to solve it - the in-falling wave condition at the horizon. Using \eqref{2.5} and \eqref{2.6} one finds 
\begin{align}
\lim_{u\rightarrow u_H}\sqrt{\frac{g_{tt}}{g_{uu}}}\, \frac{\omega}{\omega^2-m^2(u)g_{tt}}\approx \frac{4\pi T}{\omega}(u_H-u)\,,
\label{4.22}
\end{align}
which implies that the asymptotic form of \eqref{4.8} near the horizon is
\begin{align}
F''-\frac{1}{u_H-u}F'+\frac{\omega^2}{16\pi^2T^2(u_H-u)^2}F=0,
\label{4.23}
\end{align}
and the solution satisfying the in-falling wave condition at the horizon is
\begin{align}
F(u\rightarrow u_H,\omega) \sim (u_H-u)^{-\frac{i\omega}{4\pi T}}.
\label{4.24}
\end{align}
Therefore, the general Ansatz for $F$ that satisfies the in-falling wave condition at the horizon is
\begin{align}
F(u,\omega)=(u_H-u)^{-\frac{i\omega}{4\pi T}}P(u,\omega),
\label{4.25}
\end{align}
with $P(u,\omega)$ being regular at the horizon. By substituting \eqref{4.22} and \eqref{4.25} into \eqref{Pi}, one finds the boundary condition for $\Pi$ at the horizon
\begin{align}
\Pi(u_H,\omega)=1.
\label{bdyPi}
\end{align}

\subsection{Infrared limit: zero DC conductivity}
\label{sec2.2}

Let us now discuss the infrared limit in which $\omega$ is much smaller than the other energy scales of the system, i.e, the mass scale $\Lambda$ of the 2-form field and the temperature $T$ of the thermal bath. In this limit, \eqref{eqPi} reduces to
\begin{align}
\Pi'-\frac{im^2(u)\sqrt{g_{uu}g_{tt}}}{\omega}\Pi^2=0,\,\,\,\textrm{for}\,\,\, \omega\ll\Lambda,\,T.
\label{asympPi}
\end{align}
The general solution of \eqref{asympPi} is given by
\begin{align}
\Pi(u;\omega\ll\Lambda,\,T)=\frac{\omega}{-C\omega+i\int^{u_H}_u d\xi\, m^2(\xi)\sqrt{g_{uu}(\xi)g_{tt}(\xi)}}.
\label{solution}
\end{align}
The in-falling condition \eqref{bdyPi} fixes the integration constant in \eqref{solution} to be $C=-1$. Then, since $m(u)$ is finite one finds that the DC conductivity necessarily vanishes
\begin{align}
\sigma_{\textrm{DC}}=\lim_{\omega\rightarrow 0}\textrm{Re}[\sigma(\omega)]=\lim_{\omega\rightarrow 0} \frac{\omega^2}{\omega^2+\left(\int^{u_H}_0 d\xi\, m^2(\xi)\sqrt{g_{uu}(\xi)g_{tt}(\xi)}\right)^2}=0.
\label{sigmaDC}
\end{align}
One can also obtain analytically that the imaginary part of the AC conductivity at low frequencies is negative
\begin{align}
\textrm{Im}[\sigma(\omega)]\biggr|_{\omega\ll\Lambda,\,T}=\frac{-\omega \int^{u_H}_0 d\tau\, m^2(\tau)\sqrt{g_{uu}(\tau)g_{tt}(\tau)}}{\omega^2+\left(\int^{u_H}_0 d\xi\, m^2(\xi)\sqrt{g_{uu}(\xi)g_{tt}(\xi)}\right)^2}\le 0,
\label{Imsigma}
\end{align}
with the equality being saturated in the limit of zero frequency.

Therefore, in our approach $\sigma_{\textrm{DC}}$ vanishes as long as the mass scale that characterizes the 2-form bulk field, $\Lambda$, is nonzero. This result displays a certain degree of universality in the sense that it holds for the general background in \eqref{2.5} using very mild assumptions for the mass function $m(u)$, namely, that $\lim_{\epsilon \to 0}m(\epsilon) \sim \epsilon^a$, with $a=1$ or $a>3/2$, and that $m(u)$ is finite in the interior of the bulk (including at the horizon). The fact that the DC conductivity vanishes even for an arbitrarily small $\Lambda$ suggests that this energy scale may be connected to the existence of some type of condensate in the bulk. In Section \ref{sec3} we argue that $\Lambda$ can be associated with the presence of a monopole condensate in the bulk. In this sense, the DC conductivity in the QFT is analogous to an order parameter that attests the presence of a magnetic monopole condensate in the bulk.

\section{Numerical results for the AC conductivity}
\label{sec2.3}

In this section we shall specify a background to numerically evaluate the AC conductivity for some simple choices of $m(u)$ that fulfill the general requirements discussed in the previous section. We take the near-horizon approximation of the non-extremal M2-brane solution of $11$-dimensional supergravity, which corresponds (modulo a 7-sphere) to an AdS$_{3+1}$-Schwarzschild black brane metric\footnote{See, for instance, the discussions around Eqs.\ (119) and (259) of \cite{petersen}.}
\begin{align}
ds^2=\frac{4U^2}{L^2}\left(-f(U)dt^2+dx^2+dy^2\right)+\frac{L^2dU^2}{4U^2f(U)}; \,\,U\in(U_H,\infty),\,\,t,x,y\in(-\infty,\infty),
\label{3.14}
\end{align}
where $L/2$ is the radius of the asymptotically AdS space (half of the radius $L$ of the 7-sphere \cite{petersen}, which we did not write explicitly above), $f(U)=1-U_H^3/U^3$, and $U_H$ is the non-extremality parameter ($U_H=0$ for the extremal solution). The boundary of the space is at $U\rightarrow\infty$ while the horizon is at $U_H$. Defining the rescaled variable
\begin{align}
u:=\frac{U_H}{U}\Rightarrow f(U)=1-\frac{U_H^3}{U^3}=1-u^3=:h(u),
\label{3.15}
\end{align}
we can rewrite (\ref{3.14}) as follows
\begin{align}
ds^2=\frac{4U_H^2}{L^2u^2}(-h(u)dt^2+dx^2+dy^2)+\frac{L^2du^2}{4u^2h(u)};\,\,u\in(0,1),\,\,t,x,y\in(-\infty,\infty),
\label{3.16}
\end{align}
where, in the new dimensionless coordinate $u$, the boundary of the space is at $u=0$ and the horizon is at $ u_H=1$. Using (\ref{2.6}) we rewrite the non-extremality parameter $U_H$ in terms of the Hawking temperature as
\begin{align}
U_H=\frac{\pi TL^2}{3},
\label{3.17}
\end{align}
and, by substituting (\ref{3.17}) into (\ref{3.16}), we obtain the final form for the AdS$_{3+1}$-Schwarzschild background used in our numerical calculations
\begin{align}
ds^2=\frac{4(\pi TL)^2}{9u^2}(-h(u)dt^2+dx^2+dy^2)+\frac{L^2du^2}{4u^2h(u)};\,\,u\in(0,1),\,\,t,x,y\in(-\infty,\infty).
\label{3.18}
\end{align}

\begin{figure}
\begin{tabular}{cc}
\includegraphics[width=0.48\textwidth]{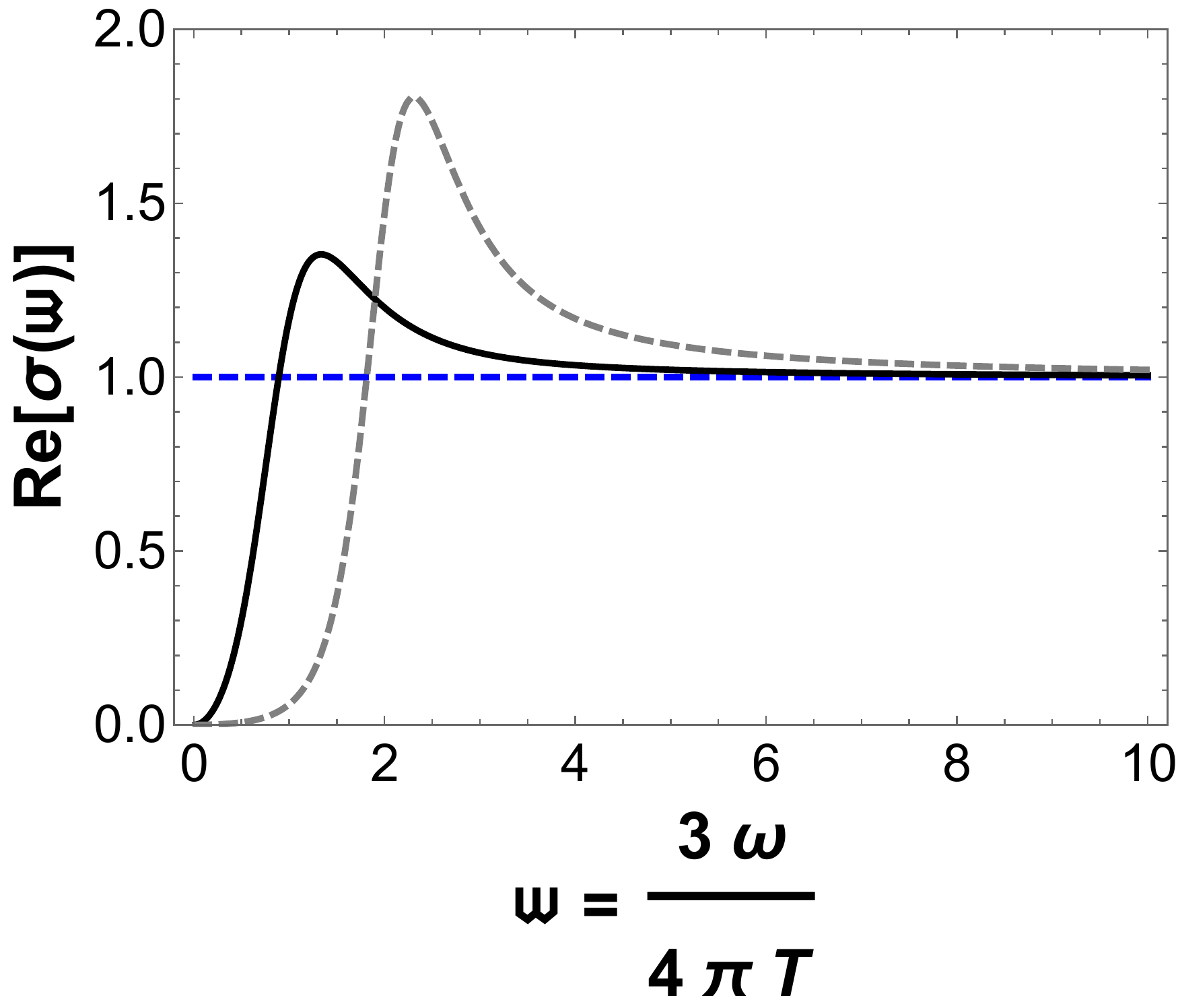} & \includegraphics[width=0.48\textwidth]{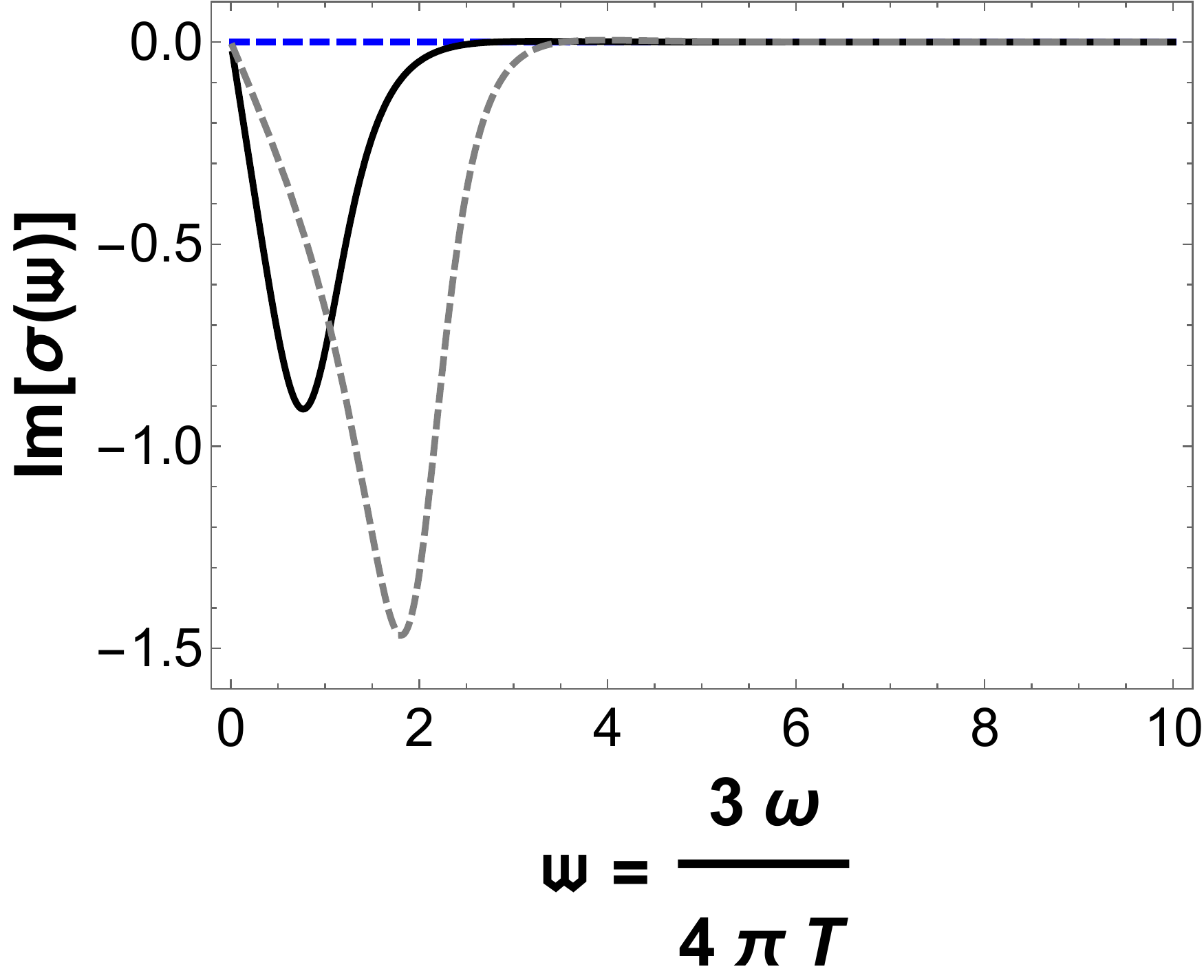}
\end{tabular}
\begin{tabular}{cc}
\includegraphics[width=0.48\textwidth]{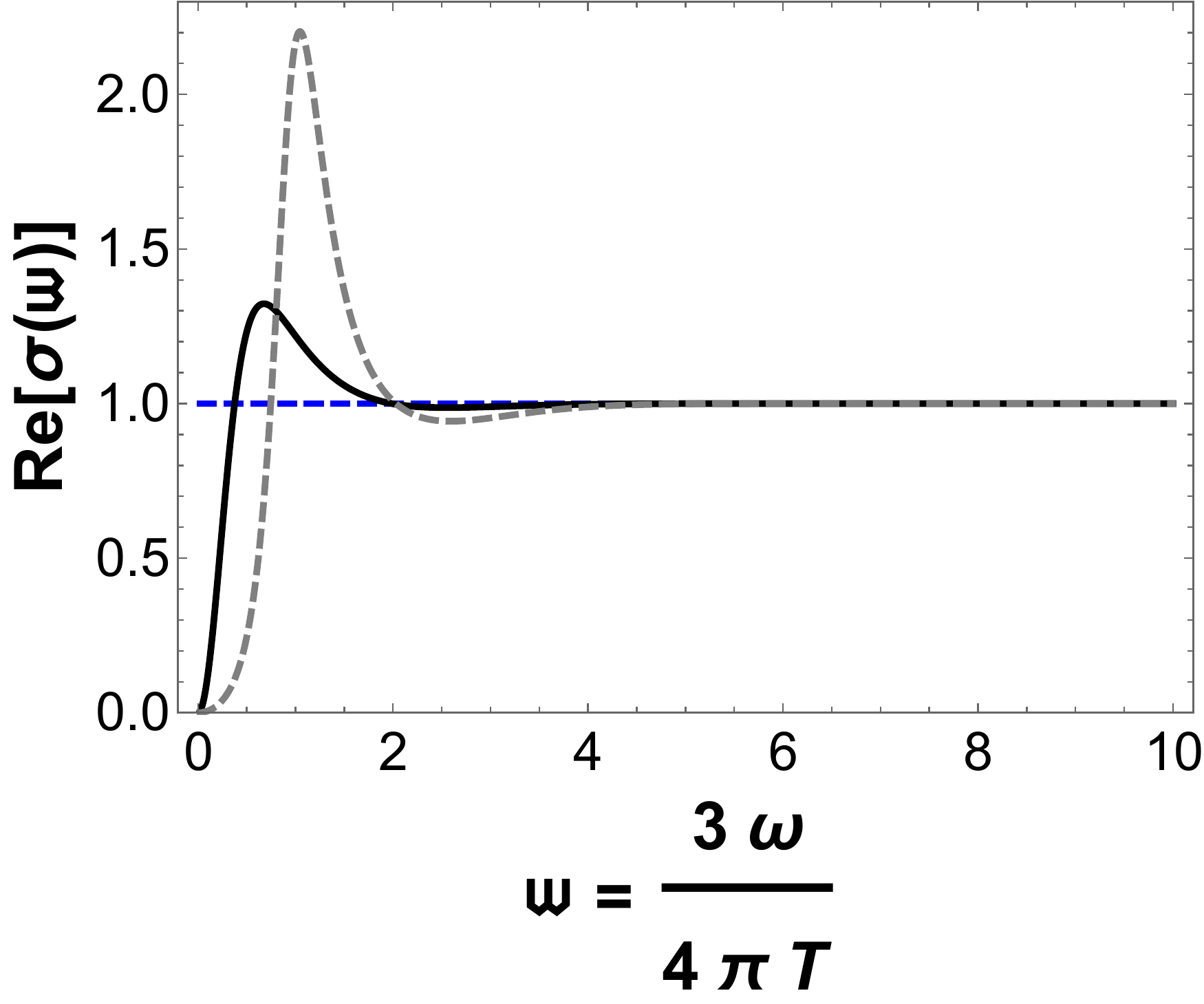} & \includegraphics[width=0.48\textwidth]{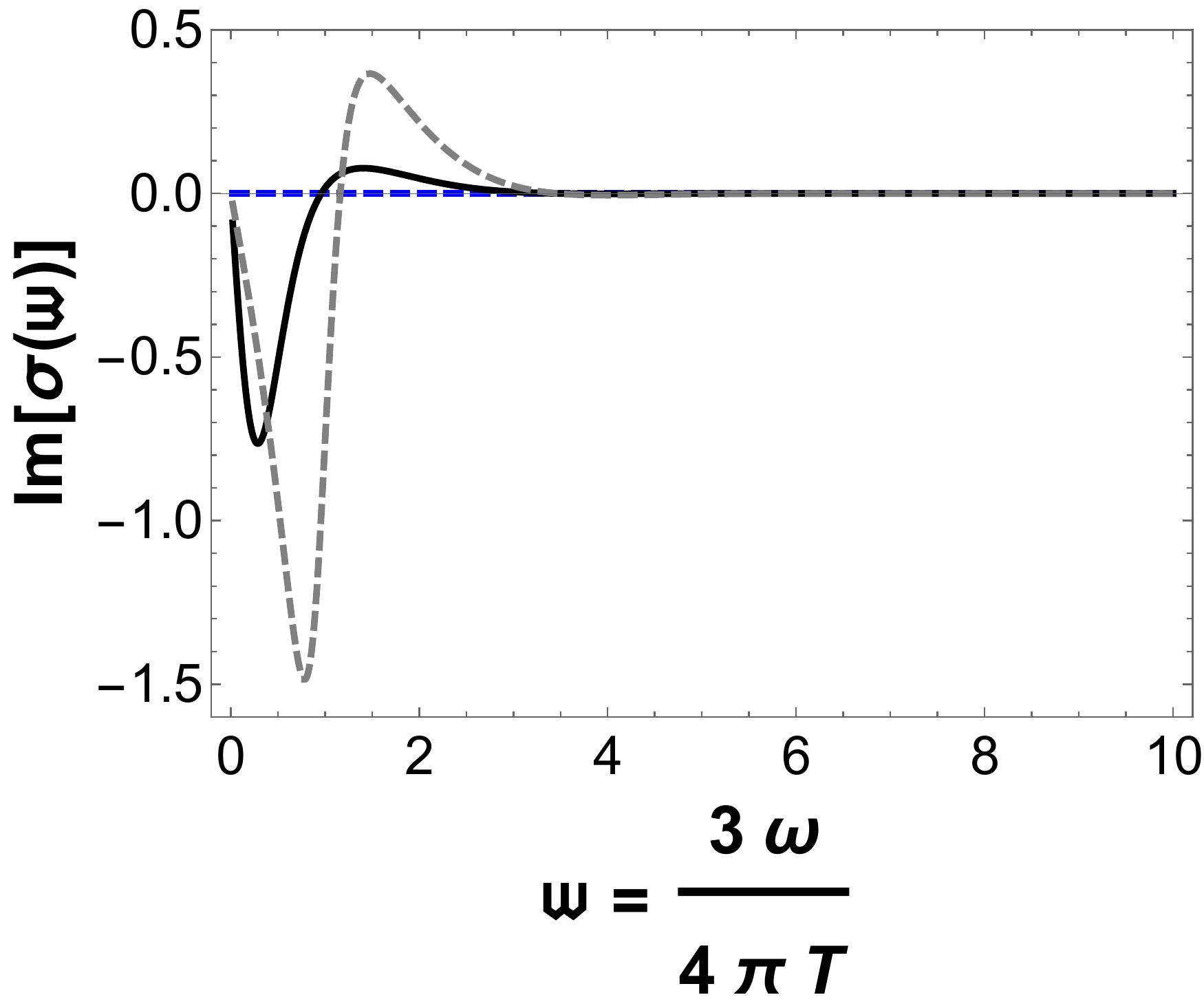}
\end{tabular}
\caption{(Color online) Real and imaginary parts of the AC conductivity as a function of the dimensionless frequency variable, $\mw=3\omega/4\pi T$, for different values of the dimensionless constant $C_\Lambda=\Lambda L/2$. The blue dashed curves correspond to the Maxwell results for the conductivity obtained by setting $C_\Lambda=0$. Solid black and gray dashed curves correspond to the results for $C_\Lambda=1$ and 2, respectively. The upper plots were generated with the mass function $M(u)=\tanh(u)$ while the lower plots were obtained with $M(u)=\tanh(u^2)$. \label{fig1}}
\end{figure}

Now, all we have to do in order to compute the real and imaginary parts of the AC conductivity \eqref{condPi} is to numerically integrate \eqref{eqPi} with the background \eqref{3.18} and impose the in-falling condition \eqref{bdyPi}. We start the integration slightly above the horizon\footnote{Note that the horizon is a singular point of \eqref{eqPi}.} and go up to an ultraviolet cutoff near the boundary. The numerical results obtained for the real and imaginary parts of the AC conductivity are shown in Fig.\ \ref{fig1} for some different choices of the mass function \eqref{2.2}.

One can see that for high frequencies the AC conductivity approaches the constant result found in the Maxwell case in \eqref{2.4}. In the opposite limit of zero frequency, both the real and imaginary parts of the conductivity vanish. For intermediate values of frequency the curves for the real and imaginary parts of the AC conductivity display extrema controlled by the value of the mass scale $\Lambda$; as one increases the value of $\Lambda$ these extrema become more pronounced and get pushed towards higher frequencies. When $\Lambda$ vanishes one recovers the constant result \eqref{2.4} for the conductivity. Therefore, the AC conductivity vanishes at zero frequency and becomes a (real-valued) constant at large frequencies, which is in agreement with the general proof given in the previous section. In the intermediate range of frequencies the results for the AC conductivity depend on the form of the effective mass function of the 2-form field and the properties of the background metric.

\section{Zero DC conductivity due to a monopole condensate in the bulk}
\label{sec3}

In this section we shall establish a physical connection between the Maxwell action and the action for a massive 2-form field in the bulk: the mass scale $\Lambda$ may be associated with the condensation of magnetic monopoles in the bulk. This is done using the well-known mechanism (originally developed for non-relativistic systems) proposed by Julia and Toulose in \cite{jt} and later generalized by Quevedo and Trugenberger \cite{qt} to describe the long wavelength excitations of a system in a phase in which a certain type of topological defects have formed a condensate. In this section we shall elaborate on how these ideas can be useful to describe the condensation of topological defects in asymptotically AdS spacetimes.

Julia and Toulose proposed in \cite{jt} a general prescription to identify the lowest lying modes of a macroscopically continuous distribution of topological defects in the context of ordered solid-state media. In this approach, the defects are described by topological currents non-minimally coupled to Abelian gauge fields. Later, Quevedo and Trugenberger \cite{qt} generalized this prescription and applied it to construct low energy effective actions describing different phases of relativistic antisymmetric tensor field theories characterized by condensates of topological defects. 

There are two general questions that one may consider in this regard. The first one concerns the conditions (i.e., values of the temperature and/or the coupling constants) under which a certain kind of topological defect may condense. The second question concerns the form of the effective field theory describing the lowest lying modes of the condensed phase, assuming that these defects have somehow condensed. This second question is the one approached in \cite{jt,qt}: one assumes that topological defects of a certain kind proliferate in spacetime until they establish a macroscopically continuous medium and works out the corresponding low energy effective field theory describing the long wavelength excitations of the condensed phase.

The condensing defects couple non-minimally to massless Abelian $p$-form gauge fields and, after the condensation process has taken place, a new massive $(p+1)$-form field emerges that describes the relevant degrees of freedom in the low energy regime of the condensed phase. Thus, the condensation of topological defects constitutes a mass gap generation mechanism whose general signature is the so-called ``rank jump phenomenon'': a massless Abelian $p$-form describing the system in the phase with diluted defects gives place to a new effective massive $(p+1)$-form describing the system in the condensed phase. Quevedo and Trugenberger refer to this as the ``Julia-Toulouse mechanism" (JTM) and, more recently, some of us generalized the JTM in various aspects and applied it to many different physical systems \cite{santiago,mcsmon,dafdc,artigao,jt-cho-su2,jt-cho-su3,mbf-m,jt-schwinger,proca-m,jt-localizacao,jt-lorentz}.

The case relevant to the present work involves pointlike Dirac magnetic monopoles \cite{dirac1,dirac2} in a $(d+1)$-dimensional bulk (the number of spatial dimensions $d$ is left unspecified in the following for the sake of generality). Such a system may be described by a Maxwell action with the Abelian 1-form gauge field non-minimally coupled to the Dirac monopoles \cite{felsager,mvf1,mvf2,ripka}, which we take to have magnetic charge $\bar{g}$
\begin{align}
S[A_\mu,j_{\mu_1\cdots\mu_{d-2}}]=-\frac{1}{4}\int_{\mathcal{M}_{d+1}}d^{d+1}x\sqrt{-g}\left(\mathcal{F}_{\mu\nu}- \bar{g}\tilde{\chi}_{\mu\nu}\right)^2,
\label{3.1}
\end{align}
where the Chern-kernel that localizes the $(d-1)$-dimensional magnetic Dirac brane $S_M^{d-1}$ is defined by the parametric equations $x^\mu=X^\mu(\lambda)$ given by
\begin{align}
\tilde{\chi}^{\mu\nu}&=\frac{1}{(d-1)!}\frac{\varepsilon^{\mu\nu\alpha_3\cdots\alpha_{d+1}}}{\sqrt{-g}}
\chi_{\alpha_3\cdots\alpha_{d+1}},\label{3.2}\\
\chi^{\alpha_1\cdots\alpha_{d-1}}&=\int_{S_M^{d-1}}d\lambda_1\cdots d\lambda_{d-1}\sum_{P(\alpha_1\cdots\alpha_{d-1})}\varepsilon_P\,\frac{\partial X^{\alpha_1}}{\partial\lambda_1}\cdots\frac{\partial X^{\alpha_{d-1}}}{\partial\lambda_{d-1}}\,\delta^{(d+1)}\left(x-X(\lambda)\right),\label{3.3}
\end{align}
where $\varepsilon_P:=+1\,(-1)$ for even (odd) permutations of $(\alpha_1\cdots\alpha_{d-1})$. The boundary of a $(d-1)$-dimensional Dirac brane is the $(d-2)$-dimensional world hypersurface of a magnetic defect in $(d+1)$-dimensions, which is localized by the monopole current density
\begin{align}
j^{\mu_1\cdots\mu_{d-2}}=\frac{1}{2}\frac{\varepsilon^{\mu_1\cdots\mu_{d-2}\alpha\beta\nu}}{\sqrt{-g}}
\nabla_\alpha\tilde{\chi}_{\beta\nu}=\nabla_\alpha\chi^{\alpha\mu_1\cdots\mu_{d-2}}.
\label{3.4}
\end{align}
This is a topological current since it is identically conserved 
\begin{align}
\nabla_\beta j^{\mu_1\cdots\beta\cdots\mu_{d-2}}=0.
\label{3.5}
\end{align}

The Maxwell action is recovered from \eqref{3.1} in the limit of completely diluted defects, i.e., when there are no monopoles in the bulk. However, what happens when there are so many magnetic monopoles in the bulk such that the system, when viewed at long distances, looks like a continuum medium of magnetic defects? What is the simplest effective theory in the bulk that is able to capture the low energy physics after the condensation of magnetic monopoles?

The gauge field $A_\mu$ in (\ref{3.1}) and its exterior derivative are both singular over the magnetic Dirac branes but the non-minimal coupling structure, $\left(\mathcal{F}_{\mu\nu}-\bar{g}\tilde{\chi}_{\mu\nu}\right)^2$, is regular on these Dirac branes and describes the observable electromagnetic fields in the presence of Dirac monopoles \cite{mvf1,mvf2,ripka}. If we assume that somehow the magnetic defects proliferate in the bulk until the establishment of a macroscopically continuous medium corresponding to the monopole condensate, the gauge field $A_\mu$ is no longer defined within the bulk and, therefore, it cannot describe the physically relevant degrees of freedom of the system in the condensed phase. 

However, notice that the non-minimal coupling is a generalized Stueckelberg-like structure with $\tilde{\chi}^{\mu\nu}$ being a Dirac delta-distribution. When the monopole currents condense, the magnetic Dirac branes occupy the entire bulk and the delta-distribution $\tilde{\chi}^{\mu\nu}$ assumes the character of a continuous field. Therefore, analogously to what happens in the Higgs mechanism, $\tilde{\chi}^{\mu\nu}$, viewed as a continuous field in the monopole condensation limit, ``eats up" the exterior derivative of the gauge field and becomes a massive 2-form field according to the following prescription \cite{jt,qt}
\begin{align}
\left(\mathcal{F}_{\mu\nu}-\bar{g}\tilde{\chi}_{\mu\nu}\right)^2 \stackrel{\textrm{cond}}{\longrightarrow}K_{\mu\nu}^2.
\label{3.6}
\end{align}
This massive 2-form field describes the long wavelength behavior of the monopole condensed phase.

An important remark regarding the prescription (\ref{3.6}) is that one observes a \emph{rank jump} of the field describing the system in passing from the diluted to the condensed phase: the massless 1-form gauge field $A_\mu$ describing the diluted phase gives place to a massive 2-form field $K_{\mu\nu}$ describing the magnetically condensed phase. In doing so, we have effectively promoted the kinetic term with magnetic defects for the $A_\mu$ field to a mass term for the $K_{\mu\nu}$ field. As mentioned above, the condensation of topological currents constitutes a type of mass generation mechanism \cite{jt,qt,artigao,mbf-m,proca-m,jt-schwinger} and the rank jump phenomenon is a general signature of this mass gap generation in the picture where the condensing currents couple non-minimally to $p$-form Abelian gauge fields \cite{jt,qt,artigao}.

In order to complete the construction of the low energy effective field theory for the magnetically condensed phase, we employ a derivative expansion for the action involving the massive field $K_{\mu\nu}$ and retain only the terms of lowest order in derivatives, which give the dominant contribution at low energies. At this point, we have two terms in our effective action, corresponding to the kinetic and mass terms for the 2-form field, which would give us a version of the action \eqref{2.1} with a constant mass for the 2-form field \cite{qt,artigao}. However, as discussed in the previous sections, we need an effective mass for this 2-form bulk field that varies with the holographic coordinate and goes to zero at the boundary in order to properly compute the holographic conductivity associated to the 2-point retarded correlation function of a conserved vector current operator at the boundary QFT sourced by the boundary value of the 2-form field. As discussed in Section \ref{sec2}, if we just take the mass of the 2-form field in \eqref{2.1} to be a radial-dependent function this violates diffeomorphism invariance. Since, in constructing effective field theories, we must preserve the physical symmetries of the system (such as diffeomorphism invariance) by insisting that the mass of the 2-form field depends on the radial coordinate we must give dynamics to this effective mass, which in turn implies that the complete effective action for the bulk monopole condensed phase should be, in general, of the form given in \eqref{diffeopreserving}, which is a diffeomorphism-invariant action.

Note that in the present scenario the mass scale $\Lambda$ in \eqref{2.2} is associated to the monopole condensate. In fact, as discussed in detail in Section 3.8 of \cite{ripka}, the complete action \eqref{diffeopreserving} may be related with the dual Abelian Higgs model describing a monopole condensate with the mass of the 2-form field, $\Lambda$, being identified with the product between the charge of the monopoles, $\bar{g}$, and the expectation value of the gauge-invariant modulus of a complex scalar field. In our action \eqref{diffeopreserving}, the field $M$ is then identified with the ratio between this modulus and its expectation value\footnote{To make clear the comparison between our action \eqref{diffeopreserving} and Eq.\ (3.143) of \cite{ripka}, the identifications we have done were the following: $\Lambda\equiv m_V = \bar{g}v = \bar{g} \langle S\rangle$ and $M\equiv S/v = S/\langle S\rangle$, where $m_V$ is the mass of the massive spin 1 particle corresponding to the lowest lying excitation of the monopole condensed phase, and $S$ is the modulus of the complex scalar field in a polar representation, $\Psi=Se^{i\bar{g}\varphi}$. The excitations associated with this scalar field describe the monopoles, which are themselves higher energy excitations. As discussed in detail in Section 3.8 of \cite{ripka}, the derivative of the phase of the complex scalar field, $\varphi$, is ``eaten up'' by the longitudinal sector of the massive dual vector field $B_\mu$, which is the electromagnetic dual of the massive 2-form field $K_{\mu\nu}$. Therefore, this phase field does not appear in the final action \eqref{diffeopreserving}. Note also that the third term in Eq.\ (3.143) of \cite{ripka} is not featured in \eqref{diffeopreserving} because we are not considering here probe charges on top of the magnetic condensate.}.

It is also important to observe that the the Maxwell field in a 4-dimensional bulk has two degrees of freedom while the number of degrees of freedom of a massive 2-form field in the same dimensionality is three, which may be easily traced back to the fact that a massive 2-form field in four dimensions is electromagnetically dual to a massive vector field \cite{ripka,artigao}. This change in the number of the degrees of freedom going from the bulk phase with diluted monopoles, described by the Maxwell action non-minimally coupled to magnetic defects, to the bulk phase with condensed monopoles (whose lowest lying excitations are described by the massive 2-form field) is associated with the mass gap generation mechanism triggered by the condensation of these monopoles. Furthermore, it is also important to point out that for any number of spacetime dimensions where a non-minimal coupling structure can be defined with respect to a 1-form gauge field\footnote{This is always possible when $d>1$, as discussed in \cite{jt-schwinger}.}, a description of the lowest lying modes of a phase characterized by a condensate of magnetically charged defects can always be given directly in terms of a massive 2-form field \cite{qt,artigao}.

Different choices for the effective mass function $m(u)$ in \eqref{2.2} correspond to parametrize in a simple way different condensation processes in the bulk, whose microscopic details are out of the scope of the JTM \cite{jt,qt}. That is, different profiles for $m(u)$ are regarded here as effective macroscopic descriptions of different monopole condensate distribution densities within the bulk. In this way, in view of the boundary condition $m(0)=0$, the monopole condensate distribution densities considered here are finite in the bulk (where the massive 2-form field describes the relevant low energy degrees of freedom) and vanish as one approaches the boundary (where the massive 2-form field gives place to the Maxwell gauge field). Thus, the general results obtained in Section \ref{sec2} can now be easily physically understood: electric field disturbances of very low frequency are sensitive to the magnetically condensed phase and the confinement of electric fluxes due to the monopole condensate makes the transport of electric charge negligible in this limit and, thus, $\sigma_{\textrm{DC}} \to 0$. On the other hand, high frequency fluctuations can resolve the magnetic condensate by probing distances smaller than $1/\Lambda$ and charge transport takes place as in the diluted phase. These properties are direct consequences of the general effective theory construction used here and, thus, we believe that they are going to be also present in any microscopic attempt to describe the process of magnetic monopole condensation in the bulk.

Also, we emphasize that the setup discussed here is related to a \emph{bulk dual superconductor} (where magnetic monopoles, instead of electric charges, are condensed) and not to a bulk usual superconductor (where electric charges condense). In the case of just a Maxwell gauge field in bulk, electric charges are in a Coulomb phase in the bulk and the associated holographic DC conductivity for the AdS$_{3+1}$-Schwarzschild background is just a constant \cite{sachdev,herzog,iqbal-liu}. When there is a magnetic monopole condensate (a dual superconducting medium) in the bulk, with the Maxwell field giving place to a massive 2-form field as discussed above, electric charges are confined in the bulk \cite{ripka,qt,artigao} and the associated holographic DC conductivity vanishes. Also, as we are going to discuss in Appendix \ref{apa}, when we have an electric condensate (a superconducting medium) in the bulk, the Maxwell field becomes Higgsed into an effective Proca field and the electric charges in the bulk are in a Higgs or screening phase such that the associated holographic DC conductivity diverges. This is the characteristic feature of a holographic superconductor \cite{HHH}. Therefore, as mentioned below Eq.\ \eqref{cond}, the different infrared results for the conductivity associated with a boundary QFT vector current operator obtained in these different calculations rely on the fact that these different pictures describe different phases of the system.

\section{Concluding remarks and perspectives}
\label{sec4}

In this work we proposed a new way to obtain a nontrivial AC holographic conductivity in a $(2+1)$-dimensional strongly coupled QFT. Our approach involves a massive 2-form field in the bulk that satisfies a special boundary condition, namely, that the effective mass of this 2-form field vanishes at the conformal boundary.  In this case, the requirement of finiteness of the action imposes that the boundary value of this 2-form field corresponds to the exterior derivative of a massless 1-form Abelian gauge field. This gauge field can be used to define the correlation function of the boundary conserved vector current in the strongly coupled thermal QFT. 

This boundary condition for the massive 2-form implies that in the ultraviolet limit of high frequencies the AC conductivity calculated in this approach reduces to the result found in the case where the source for the QFT vector current is the boundary value of a massless 1-form gauge field that exists throughout the bulk. However, at intermediate values of the frequency (when compared to the mass scale of the massive 2-form bulk field and the background temperature) the behavior of the AC conductivity obtained here differs considerably from that obtained using the Maxwell action in the bulk. In fact, it displays nontrivial profiles for its real and imaginary parts as functions of the mass of the bulk 2-form field and the dimensionless ratio between the frequency of an externally applied electric field at the boundary QFT and the temperature of the thermal bath. Most interestingly, the DC conductivity exactly vanishes. The absence of charge transport in this system is valid for a large class of black brane metrics and choices for the effective mass function $m(u)$ of the 2-form in the bulk. We also argued that this massive 2-form can be naturally understood as an effective field describing the long wavelength excitations of a condensate of Dirac magnetic monopoles in the asymptotically AdS bulk.  

Recently, there have been several studies concerning magnetic defects in holography, see for instance \cite{sachdev-cond,faulkner-iqbal,horowitz-iqbal,iqbal}, and we are currently working on establishing a more direct connection between our results and those discussed in these works. In this work, we focused on the low energy effective theory in the AdS bulk after the condensation of magnetic monopoles has taken place instead of trying to describe the condensation process per se, as recently investigated in \cite{iqbal}. The effective theory we used can be easily constructed using a simple generalization of the well-known Julia-Toulose approach to describe the condensation of topological defects in the case where the defects condense in an asymptotically AdS spacetime. Our results computed using the bulk low energy effective theory give support to the idea that a magnetic monopole condensate in the bulk leads to a vanishing DC conductivity in strongly coupled 3-dimensional QFT's \cite{sachdev-cond,faulkner-iqbal,iqbal}.

In this paper we worked in the probe approximation where the background was fixed and it would be interesting to generalize our results to the case of a dynamical background, which is a task we postpone for a future work. We also expect to obtain soon results for a nontrivial Hall conductivity calculated through an extension of the holographic setup proposed here, which takes into account the presence of topological terms in the bulk action.

\acknowledgments

The authors thank Funda\c c\~ao de Amparo \`a Pesquisa do Estado de S\~ao Paulo (FAPESP) and Conselho Nacional de Desenvolvimento Cient\'ifico e Tecnol\'ogico (CNPq) for financial support. D.R.~Granado is grateful for a PDSE scholarship from Coordena\c c\~ao de Aperfei\c coamento de Pessoal de N\'ivel Superior (CAPES). J.~Noronha thanks C.~Nu\~nez and D.~Trancanelli for discussions about the condensation of topological defects in holography. C.~Wotzasek would like to thank Prof. Dirk Rischke for a kind invitation to visit ITP-Frankfurt.

\appendix

\section{Proca conductivity}
\label{apa}

In this Appendix we derive the holographic conductivity associated with an effective Proca theory in the bulk following the same general steps discussed in the previous Sections. In other words, the effective mass of the Proca field vanishes at the boundary where it becomes a Maxwell gauge field that sources a conserved vector current in the boundary QFT. We are going to discuss also how this may be related to the holographic superconductor\footnote{Strictly speaking, the setup proposed in \cite{HHH} corresponds to a holographic superfluid or to a holographic superconductor in the limit of a nondynamical boundary electromagnetic gauge field. The absence of a dynamical gauge field at the boundary in such description is tied to the chosen boundary condition for the bulk gauge field, namely, the Dirichlet boundary condition. As proposed in \cite{salvio1,salvio2,salvio3,salvio4}, a dynamical gauge field at the boundary may be turned on by using a Neumann boundary condition for the bulk gauge field, allowing the study of the important role played by dynamical gauge fields in holographic superconductors.} setup proposed in \cite{HHH}.

In analogy to the discussion carried out in Section \ref{sec2}, let us begin by writing down a bulk action for a quadratic massive vector field, $A_\mu$, interacting with a real scalar field, $m$, in an asymptotically AdS$_{3+1}$ spacetime
\begin{align}
S_{\textrm{Proca+scalar}}=-\int_{\mathcal{M}_{3+1}}d^{4}x\sqrt{-g} &\left[\frac{1}{4}\mathcal{F}_{\mu\nu}^2+\frac{m^2}{2}A_\mu^2+\frac{1}{2}\left(\nabla_\mu m\right)^2 +V(m)\right].
\label{diffeopreserving2}
\end{align}
This action is equivalent to the action for the Maxwell-complex scalar field theory in Eq. (4) of \cite{HHH}, with the Maxwell gauge field becoming Higgsed into an effective Proca field after its longitudinal sector became a physical degree of freedom by ``eating up'' the derivative of the phase field, $\varphi$, of the complex scalar field $\Psi\equiv me^{i\varphi}$ describing the electric condensate. The sector of the complete action \eqref{diffeopreserving2} which depends on the massive vector field is given by the following Proca action, where as before, we are taking an Ansatz for the effective mass function which depends only on the radial coordinate
\begin{align}
S=-\int_{\mathcal{M}_{3+1}}d^4x\sqrt{-g}\left[\frac{1}{4}\mathcal{F}_{\mu\nu}^2+\frac{m^2(u)}{2}A_\mu^2\right].
\label{a1}
\end{align}
As remarked in Section \ref{sec2}, we note that the partial action \eqref{a1} does not have a covariantly conserved energy-momentum tensor because it neglects the dynamics of the effective mass field (taken into account in the complete theory \eqref{diffeopreserving2}, which does have a covariantly conserved energy-momentum tensor). However, for the calculation of 2-point correlation functions of components of the Proca field, the results obtained by using the simpler partial action \eqref{a1} with some prescribed profile for $m(u)$ which vanishes at the boundary may be in principle also obtained from the complete action \eqref{diffeopreserving2} by choosing an adequate potential for the (relevant) effective mass field. Note also that when this mass vanishes the electric condensate in the bulk disappears and we recover from both, \eqref{diffeopreserving2} and \eqref{a1}, the Maxwell action.

In order to calculate the conductivity we only need the equation of motion for the $x$-component of the Proca field, which decouples from the other components and may be written in the form below
\begin{align}
\partial_u\left(\sqrt{\frac{g_{tt}}{g_{uu}}}A_x'\right)+\sqrt{\frac{g_{uu}}{g_{tt}}}\left(\omega^2-m^2(u)g_{tt}\right)A_x=0.
\label{a2}
\end{align}

Let us now define the Ansatz
\begin{align}
A_x(u,\omega)=A_x^0(\omega)F(u,\omega),
\label{a3}
\end{align}
such that, in terms of $F$, the Dirichlet boundary condition for the Proca field is given by $F(0,\omega)=1$. The sector of the on-shell boundary action that contributes to the retarded Proca propagator $G_{xx}^R(\omega)$ is given by
\begin{align}
S^{\textrm{boundary}}_{\textrm{on-shell}}=-\frac{1}{2}\int \frac{d\omega d^2\v{q}}{(2\pi)^3}A_x^0\,^*(\omega,\v{q})\left[-\lim_{\epsilon\rightarrow 0}\sqrt{\frac{g_{tt}(\epsilon)}{g_{uu}(\epsilon)}}\, F'(\epsilon,\omega,\v{q})\right]_{\textrm{on-shell}}^{\textrm{in-falling}} A_x^0(\omega,\v{q}) + (\cdots\,\!),
\label{a4}
\end{align}
where $(\cdots\,\!)$ denotes terms that do not contribute to $G^R_{xx}(\omega)$. From \eqref{a4}, we immediately read off the formula for the AC conductivity
\begin{align}
\sigma(\omega)=\frac{iG_{xx}^R(\omega)}{\omega}=-\lim_{\epsilon\rightarrow 0}\sqrt{\frac{g_{tt}(\epsilon)}{g_{uu}(\epsilon)}} \frac{iF'(\epsilon,\omega)}{\omega}\biggr|_{\textrm{on-shell}}^{\textrm{in-falling}}.
\label{a5}
\end{align}

\begin{figure}
\begin{tabular}{cc}
\includegraphics[width=0.48\textwidth]{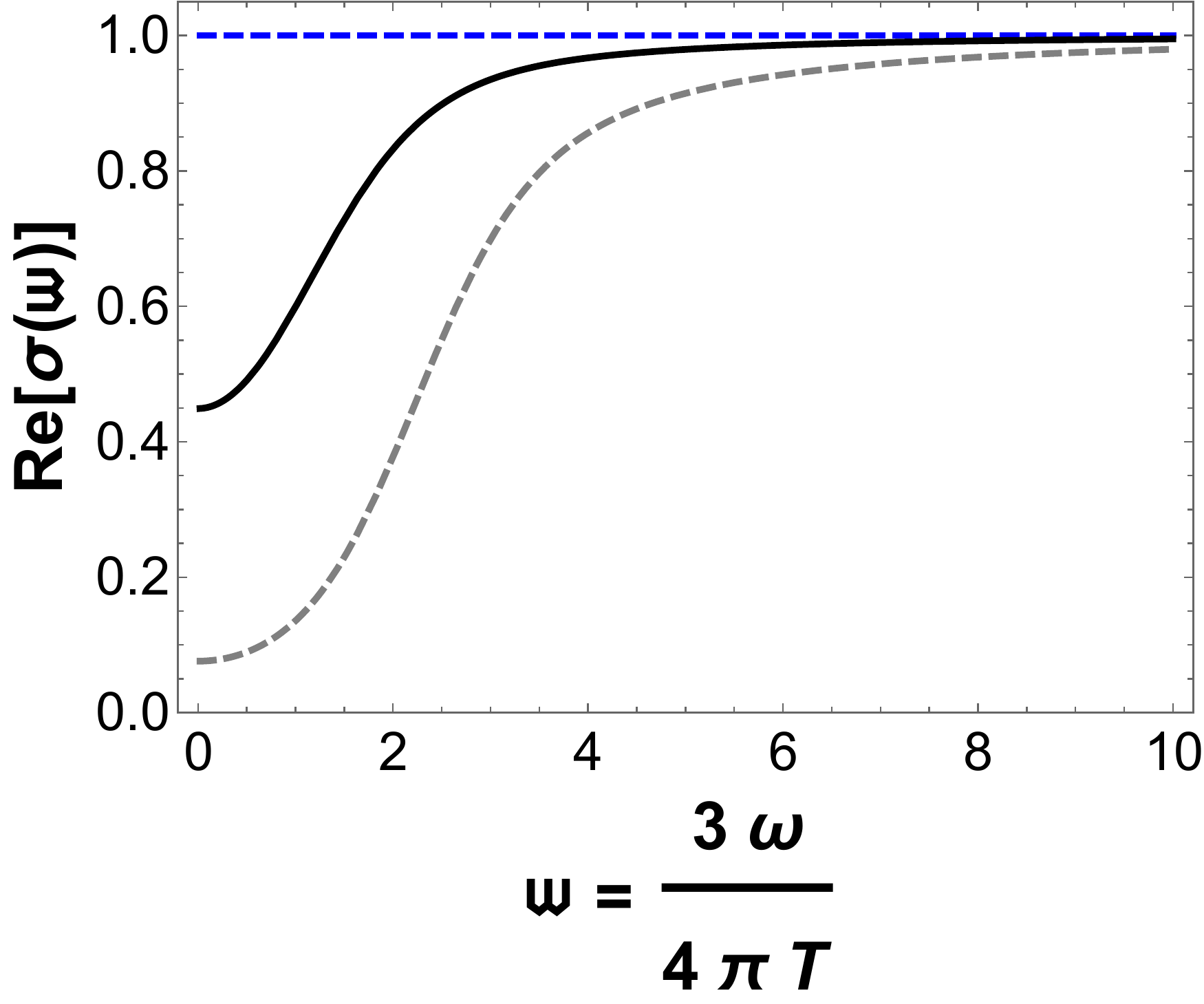} & \includegraphics[width=0.48\textwidth]{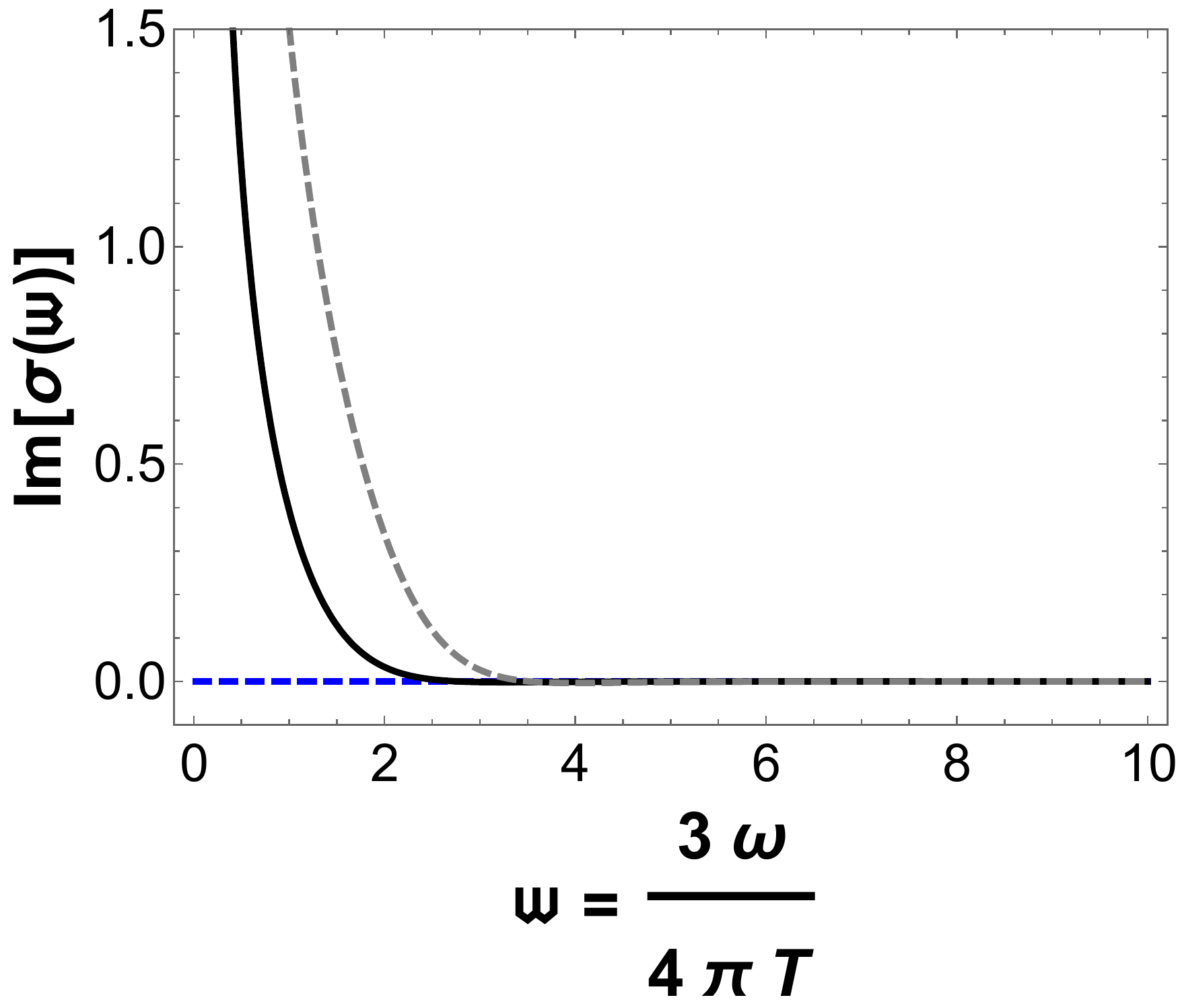}
\end{tabular}
\begin{tabular}{cc}
\includegraphics[width=0.48\textwidth]{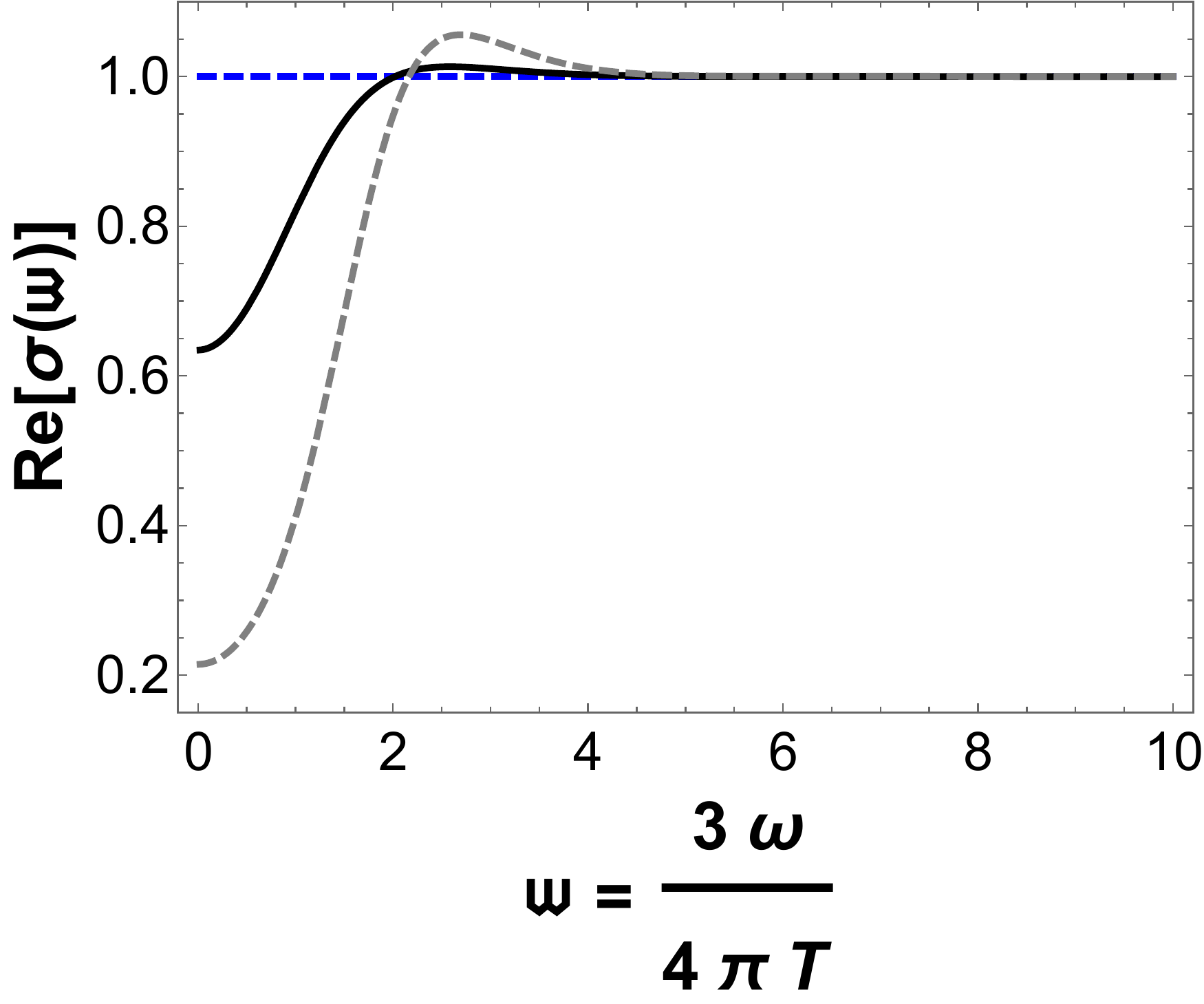} & \includegraphics[width=0.48\textwidth]{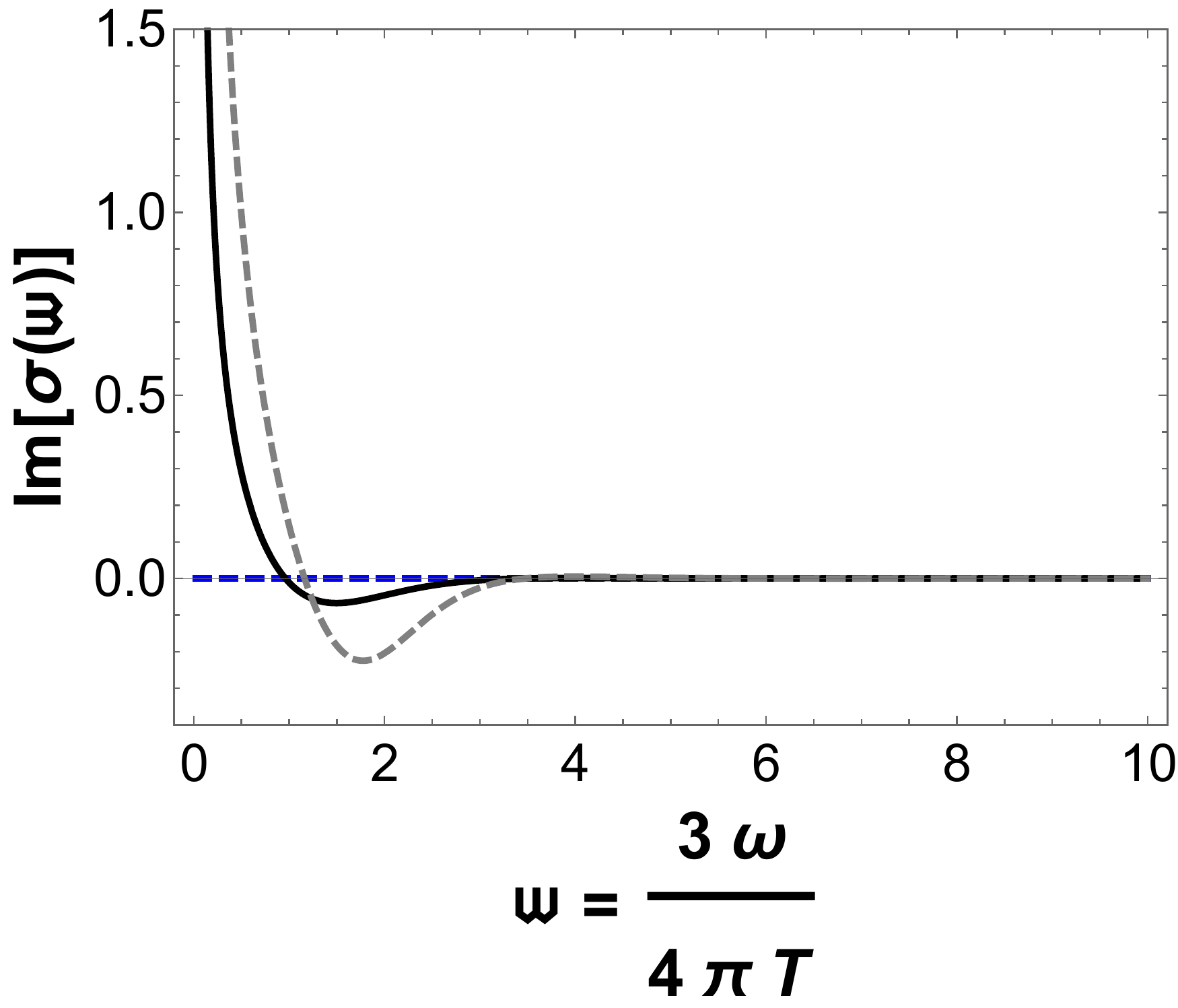}
\end{tabular}
\caption{(Color online) Real and imaginary parts of the AC conductivity as a function of the dimensionless frequency variable, $\mw=3\omega/4\pi T$, for different values of the dimensionless constant $C_\Lambda=\Lambda L/2$. The blue dashed curves correspond to the Maxwell results for the conductivity obtained by setting $C_\Lambda=0$. Full black and gray dashed curves correspond to the results for $C_\Lambda=1$ and 2, respectively. The upper plots were generated with the mass function $M(u)=\tanh(u)$, while the lower plots were obtained with $M(u)=\tanh(u^2)$. \label{fig2}}
\end{figure}

Now, we define the following quantity
\begin{align}
\Pi(u,\omega)=-\sqrt{\frac{g_{tt}}{g_{uu}}}\frac{iF'(u,\omega)}{\omega F(u,\omega)},
\label{a6}
\end{align}
and follow the same steps as before to show that the equation of motion \eqref{a2}, rewritten in terms of $\Pi$, is given by
\begin{align}
\Pi'+i\omega\sqrt{\frac{g_{uu}}{g_{tt}}}\left[\Pi^2-\frac{\omega^2-m^2(u)g_{tt}}{\omega^2}\right]=0,
\label{a7}
\end{align}
and that $\Pi(u_H)=1$, with the conductivity given by the boundary value of $\Pi$. At low frequencies, one can then show from \eqref{a7} that the imaginary part of the Proca conductivity diverges as $\mathcal{O}(\omega^{-1})$ which, due to the Kramers-Kronig relations, implies that the real part of the Proca conductivity displays a delta distribution at zero frequency\footnote{This delta distribution for the real part of the conductivity at zero frequency is hard to see numerically and it is inferred from the divergent behavior of the imaginary part of the conductivity at zero frequency.} \cite{HHH}, that is, the Proca DC conductivity diverges. The numerical results for the Proca AC conductivity for the same choices of the mass function \eqref{2.2} and background \eqref{3.18} used in Section \ref{sec2.3} are shown in Fig.\ \ref{fig2}.

These results are clearly different from those derived in Section \ref{sec2.3}, which are displayed in Fig.\ \ref{fig1}, with the main difference relying on the fact that the DC conductivity associated with the massive 2-form field vanishes while the DC Proca conductivity diverges. In fact, the curves obtained in Fig.\ \ref{fig2} for the Proca conductivity are qualitatively the same as the curves shown in Fig.\ 2 of \cite{HHH}, which were obtained in the context of a probe holographic superconductor.

Indeed, by employing the AdS$_{3+1}$-Schwarzschild metric written in the form of Eq.\ (1) of \cite{HHH}, the Proca equation of motion \eqref{a2} is equivalent to the equation of motion (13) of \cite{HHH} for a Maxwell perturbation in the probe holographic superconductor described by the Maxwell-complex scalar field theory defined on a fixed background corresponding to the AdS$_{3+1}$-Schwarzschild space, provided we identify the effective radial-dependent Proca mass $m(u)$ with the modulus of the complex scalar field in the context of the probe holographic superconductor discussed in \cite{HHH}. This scalar field has two normalizable modes near the boundary which act as sources for two different charged scalar operators at the boundary QFT with different scaling dimensions. Then, one may impose two different Dirichlet boundary conditions where one of these modes vanishes at the boundary while the other one remains finite. According to Eq.\ (10) of \cite{HHH}, when the boundary charged scalar operator with dimension 1 condenses, the expectation value of the boundary charged scalar operator with dimension 2 vanishes, and vice-versa. 

From the identification proposed above and also by comparing the curves in Fig.\ \ref{fig2} with the curves in Fig.\ 2 of \cite{HHH} we see that the boundary condition we imposed, namely, that the effective mass of the Proca field vanishes at the boundary, should be related to the condensation of a charged scalar operator at the boundary and that the dimension of such operator is controlled by the near-boundary asymptotics of the mass function for the Proca field\footnote{Note that for the mass functions used to obtain the plots in Fig.\ \ref{fig2} we have near the boundary $m(\epsilon)\sim\epsilon$ and $m(\epsilon)\sim\epsilon^2$ for the upper and lower plots, respectively.}.

\end{document}